\documentclass[prb,amsmath,amssymb,reprint]{revtex4-2}
\usepackage{amsmath}
\usepackage{graphicx}
\usepackage{bm}
\usepackage{xr}
\usepackage{chemformula}
\usepackage{amsfonts}
\usepackage{times}
\usepackage{helvet} 
\usepackage{courier} 
\usepackage{url}
\usepackage{placeins}

\newcommand{\DefineFigure}[6]{%
  \expandafter\newcommand\csname #1\endcsname{%
    \begin{figure}[#2]
      \centering
      \includegraphics[width=\linewidth]{#4}
      \caption{#5}
      \label{fig:#6}
    \end{figure}
  }%
}

\newcommand{\DefineMultiFigure}[6]{%
  \expandafter\newcommand\csname #1\endcsname{%
    \begin{figure}[#2]
      \centering
      \includegraphics[width=0.8\linewidth]{#3}
      \includegraphics[width=0.9\linewidth]{#4}
      \caption{#5}
      \label{fig:#6}
    \end{figure}
  }%
}

\newcommand{\DefineWideFigure}[6]{%
  \expandafter\newcommand\csname #1\endcsname{%
    \begin{figure*}[#2]
      \centering
      \includegraphics[width=#3]{#4}
      \caption{#5}
      \label{fig:#6}
    \end{figure*}
  }%
}

\DefineWideFigure{FigureOne}{!t}{0.95\textwidth}{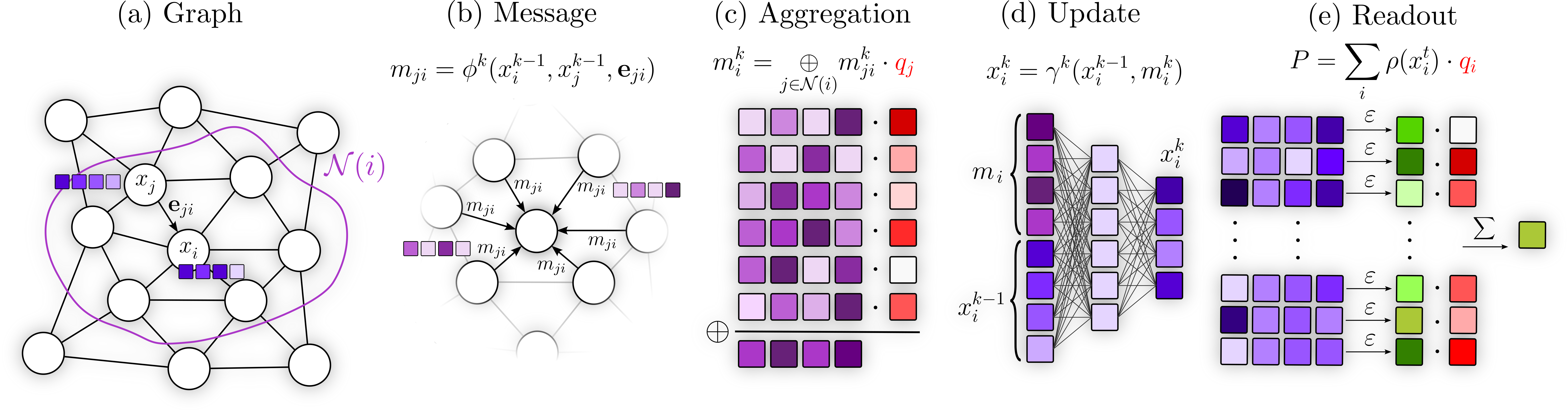}{Illustration of the different steps involved in message-passing. (a) Graph construction and 
    neighbourhood definition with two node embeddings shown. (b) Message passing from neighbouring nodes with two messages shown. (c) Aggregation of messages 
    to compute the full message, with adaptations to account for node existence shown in red. (d) Node embedding update based on the message and the previous embedding. (e) Readout layer 
    using node predictions to compute a total property modified to take into account the existence of nodes.}{fig1_gnn_illustration}

\DefineWideFigure{FigureTwo}{}{0.8\textwidth}{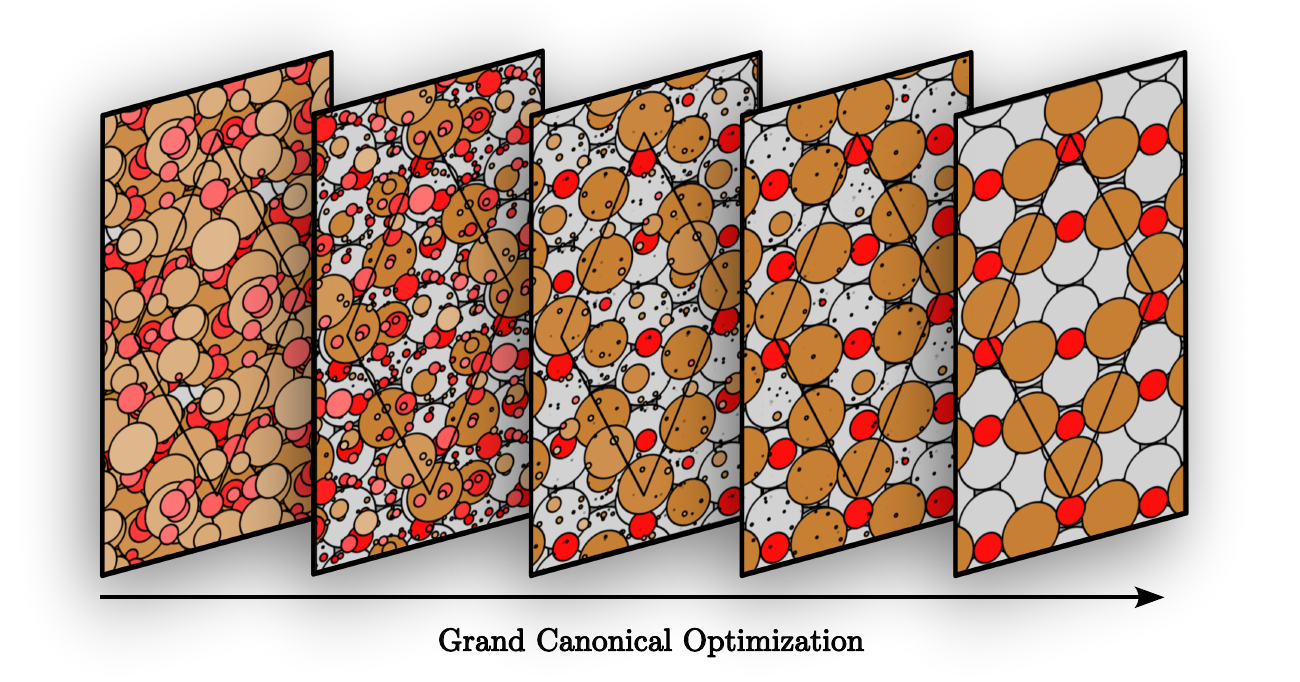}{Configuration snapshots from grand canonical optimization trajectory. The area of the atoms is linearly proportional to their existence and lighter colors indicate atoms 
    that are further from the surface slab for which the copper atoms are depicted in gray. Initially, the configurations involve many atoms of various existences inserted at random and, as the optimization
    progresses only those atoms that provide stability at the chosen chemical potentials survive with the rest going out of existence.}{opt_frames}

\DefineWideFigure{FigureThree}{}{0.8\textwidth}{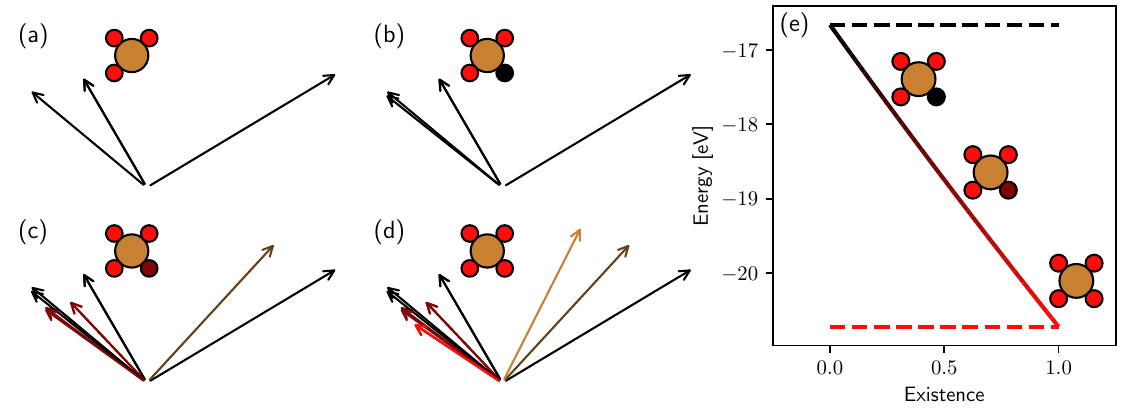}{The conditions on the inclusion of existence in a graph neural network, tested 
    with the trained network described in Section \ref{sec:training_dataset}.
    In (a) two dimensions of the learned atomic embeddings are plotted for a \ch{CuO3} cluster which can 
    be compared to (b) where the same embeddings are shown for the same configuration with an additional oxygen atom with null existence, 
    this confirms that embedding equivalence is upheld. In (c) the existence of the additional oxygen atom is increased to 1/2 and in (d) 
    it is increased 1, this shows how the embeddings smoothly sweep across the embedding space as a function of the existence. 
    Finally in (e) both read-out equivalence and continuity are tested, showing 
    the energy as a function of the existence of an atom in a small cluster along side the energy 
    of the cluster without and with the extra atom.}{existence_conditions}

\DefineMultiFigure{FigureFour}{}{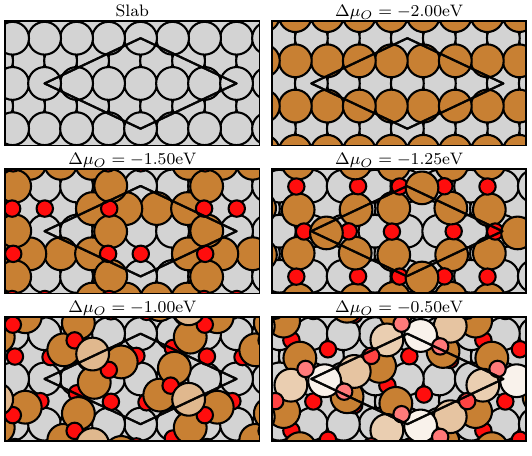}{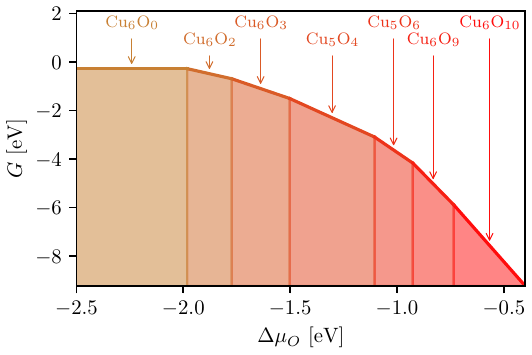}{Left: The $c(6\times 2)$ copper slab along with the most favourable configuration found at the five different chemical potentials 
    targeted with the grand canonical optimization scheme. Right: Gibbs free energy diagram calculated using the model and the configurations 
    generated by the grand canonical optimization procedure. The small discrepancy of the Gibbs free energy of the \ch{Cu6O0} configuration, not being equal to zero, is 
    caused by a slight relaxation of by the added layer compared to the bare slab.}{cu_6x2_configs}

\DefineWideFigure{FigureFive}{}{0.9\textwidth}{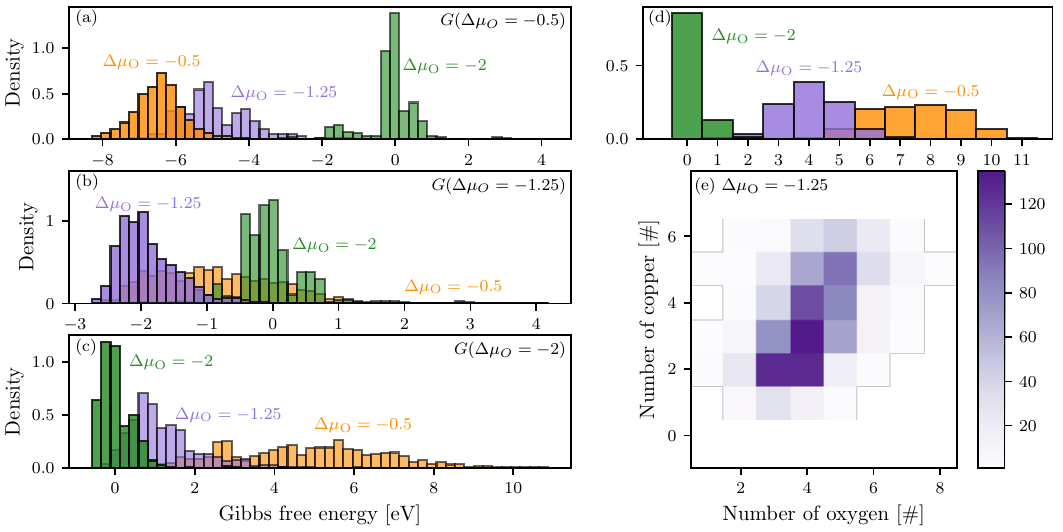}{(a-c): Distributions of Gibbs free energies calculated at different chemical potentials (in eV), each color corresponds to 
    configurations identified by our optimization procedure with a specific chemical potential for oxygen. In each case the configurations 
    that were generated at the desired chemical potential have lower Gibbs energies, showings the methods capability at targeting specific conditions. 
    (d): Distributions of the number of oxygen in configurations generated at different chemical potentials. (e): The number of oxygen and copper
    in configurations generated at $\Delta \mu_\mathrm{O} = -1.25$ eV.}{gibbs_dist}

\DefineMultiFigure{FigureSix}{}{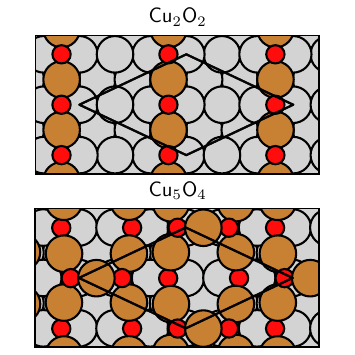}{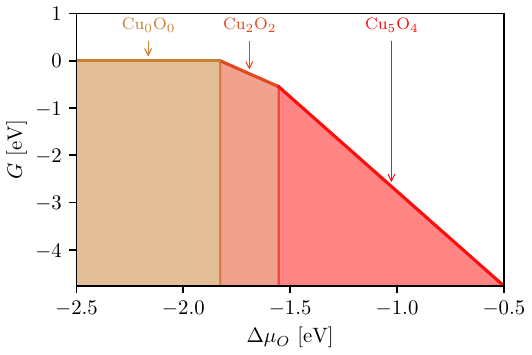}{Left: The two stable oxide phases of 
    the surface at the DFT level. Right: Gibbs free energy diagram as a function of the oxygen
    chemical potential at the DFT level, identifying \ch{Cu5O4} as the preferred phase 
    across a range of chemical potentials.}{dft_gibbs_diagram}

\begin{document}
\newcommand{\zenodo}{\url{https://doi.org/10.5281/zenodo.16420059}}
\newcommand{\gitlab}{\url{https://gitlab.com/Mads-Peter/quail}}

\author{Mads-Peter Verner Christiansen}
\author{Bjørk Hammer}
\email{hammer@phys.au.dk}
\affiliation{Center for Interstellar Catalysis, Department of Physics and Astronomy, 
    Aarhus University, DK-8000 Aarhus C, Denmark}

\title{Gradient-based grand canonical optimization enabled by graph neural networks with fractional atomic existence.}

\begin{abstract}
Machine learning interatomic potentials have become an indispensable tool for materials science, enabling the study of 
larger systems and longer timescales. State-of-the-art models are generally graph neural networks that employ 
message passing to iteratively update atomic embeddings that are ultimately used for predicting properties. In this work 
we extend the message passing formalism with the inclusion of a continuous variable that accounts for fractional atomic existence. 
This allows us to calculate the gradient of the Gibbs free energy with respect to both the Cartesian coordinates of atoms and their existence. 
Using this we propose a gradient-based grand canonical optimization method and document its capabilities for a Cu(110) surface 
oxide.
\end{abstract}

\maketitle

\section{Introduction}

The potential energy surface (PES) is a central concept in computational materials science at the 
atomic scale. Over the last two decades advancements in machine learning techniques have 
enabled the training of machine learning interatomic potentials (MLIPs) that can serve as a reliable substitute 
for quantum mechanical calculations. These MLIPs offer accuracy comparable to that of their training data at a 
fraction of the cost and with much more favourable scaling for larger systems~\cite{Behler2007,Bartok2010}. 
Additionally, they open avenues for exploring the PES that are not feasible in their quantum mechanical counterparts, 
in this paper we explore one such avenue - namely the incorporation of atomic existence as a continuous variable 
on equal footing to the Cartesian coordinates. 

State-of-the-art models are generally graph neural networks, which can 
efficiently encode and utilize invariant~\cite{schuttSchNetContinuousfilterConvolutional2017, gilmerNeuralMessagePassing2017, xieCrystalGraphConvolutional2018, unkePhysNetNeuralNetwork2019} 
and equivariant~\cite{thomasTensorFieldNetworks2018, gasteigerDirectionalMessagePassing2019, andersonCormorantCovariantMolecular2019, schuttEquivariantMessagePassing2021, batznerEquivariantGraphNeural2022, batatiaMACEHigherOrder2023} 
properties to learn the PES. These advancements, alongside the production and curation of large high quality datasets for molecules~\cite{ghahremanpour2018alexandria}, 
two-dimensional materials \cite{mounetTwodimensionalMaterialsHighthroughput2018}, crystals \cite{dengCHGNetPretrainedUniversal2023, pengOpenLAMChallenges2025}, 
nanoclusters~\cite{mannaDatabaseLowenergyAtomically2023} and inorganic materials~\cite{barroso-luqueOpenMaterials20242024},
have enabled the development of universal machine learning potentials ($u$MLIPs) underpinning the importance 
of graph neural networks for computational materials science~\cite{merchantScalingDeepLearning2023, batatiaFoundationModelAtomistic2024, parkScalableParallelAlgorithm2024, neumannOrbFastScalable2024, kimDataefficientMultifidelityTraining2024, yangMatterSimDeepLearning2024, bochkarevGraphAtomicCluster2024, fuLearningSmoothExpressive2025}.  

In parallel to the development of these descriptions of the PES, methods to explore it 
have also been the subject of much work e.g. in the realm of surfaces and interfaces~\cite{hormannMachineLearningDatadriven2025}. 
This involves optimization algorithms for locating both the lowest energy minima, algorithms for locating transition states and increasingly 
data-driven generative machine learning models. Quintessential methods within the last two categories include random structure search, basin-hopping, simulated annealing and the nudged-elastic band method \cite{kirkpatrickOptimizationSimulatedAnnealing1983, walesGlobalOptimizationBasinHopping1997a, jonssonNudgedElasticBand1998, pickardInitioRandomStructure2011}. 
Beyond these, families of methods have been reported on in the literature, such as 
evolutionary algorithms~\cite{l.johnstonEvolvingBetterNanoparticles2003, Kolsbjerg2018, Behler2020, arrigoniEvolutionaryComputingMachine2021, fallsXtalOptEvolutionaryAlgorithm2021, wangMAGUSMachineLearning2023, wanzenbockClinamen2FunctionalstyleEvolutionary2024}, 
Bayesian optimization~\cite{hernandez-lobatoParallelDistributedThompson2017, yamashitaCrystalStructurePrediction2018, Todorovic2019, Bisbo2020, kaappaGlobalOptimizationAtomic2021, lyngbyBayesianOptimizationAtomic2024}, 
particle swarm optimization~\cite{callGlobalMinimumStructure2007, lvParticleswarmStructurePrediction2012, wangCALYPSOStructurePrediction2016}.
In addition to these families, a number of novel exploration strategies have been proposed using various techniques such as reinforcement learning~\cite{simmSymmetryAwareActorCritic3D2020, meldgaardGeneratingStableMolecules2021a, millsExploringPotentialEnergy2022}, symmetry~\cite{Brix2024Cascading}, graph classification~\cite{slavenskyAcceleratingStructureSearch2024}, distribution functions~\cite{taylorRAFFLEActiveLearning2025}, procreation operators~\cite{lysgaardGeneticAlgorithmProcreation2014} or ephemeral potentials~\cite{j.pickardTheorydrivenDiscoveryIntroducing2025}.
Finally, various ways of exploiting MLIPs to enhance performance have been proposed~\cite{ouyangGlobalMinimizationGold2015, tongAcceleratingCALYPSOStructure2018, timmermannMathrmIrOSurface2020, christiansenAtomisticGlobalOptimization2022a}.

A prominent approach is some form of transformation, alteration, modification or approximation of the PES that eases the 
task of exploring it. In particular, in basin-hopping the continuous PES is transformed 
by local optimization into a piece-wise constant function aiding in the task of global optimization.
In nudged-elastic band calculations, intermediate images along the reaction path are 
optimized in a modified PES that also includes a spring-force between adjacent images. 
Likewise, the inclusion of uncertainty may be considered such an alteration as is done in Bayesian techniques and 
has seen use as an aid for exploration in molecular dynamics \cite{vandaeroordHyperactiveLearningDatadriven2023, kulichenkoUncertaintydrivenDynamicsActive2023, zaverkinUncertaintybiasedMolecularDynamics2024}.
Other recent examples, include work on so-called complementary energy landscapes, where a complementary function based 
on atomic feature similarity is used to propose new configurations \cite{slavenskyGeneratingCandidatesGlobal2023}. 
Similarly, a modified PES may be constructed that favours high symmetry configurations increasing exploration of 
those regions of the PES \cite{huberTargetingHighSymmetry2023}. In hyperspatial optimization, additional dimensions are introduced allowing gradient-based optimization to move atoms through regions otherwise 
forbidden due to the strong repulsion at short distances, again aiding in finding low energy configurations \cite{pickardHyperspatialOptimizationStructures2019}. 
Others have introduced methods that enable the interpolation between chemical elements, allowing gradient-based optimization 
to swap the chemical identity of atoms in trial configurations \cite{kaappaAtomicStructureOptimization2021}. 
Another modification is the concept of fractional or continuous atomic existence, where the 
existence of an atom is introduced as an additional variable in the construction of a surrogate potential, 
allowing atoms to not only move around in Cartesian space but also to whip in and out of existence,
again allowing local optimization pathways that are otherwise forbidden by the unmodified PES \cite{larsenMachinelearningenabledOptimizationAtomic2023, larsenGlobalAtomicStructure2025}. 

\FigureOne

The work on fractional existence by Larsen et al. \cite{larsenMachinelearningenabledOptimizationAtomic2023,larsenGlobalAtomicStructure2025} employed Gaussian Process regression based on global descriptors, 
which imposed the need for constraining the total existence to a fixed value and limited the training data 
to configurations with a particular number of atoms. 
In this paper we extend the concept of continuous atomic existence to graph neural networks, allowing 
its use with state-of-the-art models with no restrictions on the dataset and allows us to lift the 
constraint on the total existence. We derive the necessary conditions for the meaningful inclusion of 
atomic existence in graph neural networks and show how a minor modification of the message-passing framework 
is sufficient to guarantee those conditions. Grand canonical optimization involves including 
the number of atoms as a degree of freedom and is a topic of wide interest~\cite{chenAutomatedSearchOptimal2022, liSituStructureMoDoped2023, couzinieMachineLearningSupported2025, zhangGOCIAGrandCanonical2025}
We use the inclusion of fractional atomic existence in a GNN to propose a gradient-based 
grand canonical optimization procedure where the Gibbs free energy is the objective. 

\section{Methods}

\subsection{Graphs \& Message-Passing}
\label{sec:graphs_mp}

Graphs are a natural structure to represent pairwise relations between objects such 
as between cities for a travelling salesman type problem, between people in a social network or 
between particles for various physical interactions. A powerful approach to machine learning 
with graphs are graph neural networks employing message passing to iteratively update node or 
edge representations. The message-passing framework allows learning of embeddings that go beyond 
simple pairwise attributes due to the iterative aggregation of information from neighboring nodes, 
allowing representations to reflect context-dependent interactions beyond direct pairwise relations.
Denoting the representation or embedding of node $i$ at iteration $k$ as $x_i^k$ and 
the edge representation of the directed edge from $j$ to $i$ as $\mathbf{e_{ji}}$ which may include vectorial elements, message passing 
may be defined as 

\begin{equation}
    \begin{aligned}
        x_i^k &= \gamma^k\left[ x_i^{k-1}, 
        \underset{j\in\mathcal{N}(i)}{\Large\oplus}
        \ \phi^k\left(x_i^{k-1}, x_j^{k-1}, \mathbf{e}_{ji}\right)\right] \\
        &= \gamma^k\left[ x_i^{k-1},
        \underset{j\in\mathcal{N}(i)}{\Large\oplus} \ m_{ji}^k \right] \\
        &= \gamma^k\left[ x_i^{k-1}, m_{i}^k \right].
    \end{aligned}
    \label{eq:message_passing}
\end{equation}

Starting from the right in Eq. \eqref{eq:message_passing} a message $m_{ji}^k$ from node $j$ to node $i$ is constructed 
through the message function $\phi^k$ using the node representations and edge representation of the involved nodes. 
Messages are aggregated by a permutation invariant aggregation operator $\oplus$ over the nodes that are in the 
neighbourhood of node $i$ denoted $\mathcal{N}(i)$, producing the aggregated message $m^k_i$. The update function 
$\gamma^k$ combines the previous representation of node $i$ with the message $m_i^k$. To enable learning 
the message $\phi^k$ and update $\gamma^k$ functions can be neural networks. We illustrate the different steps of message passing in Figure \ref{fig:fig1_gnn_illustration}.

After a predetermined number of message passing iterations $t$ the node embeddings may be 
used to predict global graph properties or local node properties, for example a local 
property $p_i$ of node $i$ may be predicted as 
\begin{equation}
    p_i = \rho(x_i^t).
\end{equation}
Where $\rho$ is a, possibly learnable, function mapping $x_i^t$ to the property $p_i$. This is 
often called a \textit{readout} layer. Likewise, global graph properties may be predicted 
as a function of an aggregation of all node features or through the aggregation of 
predicted node properties, the latter case can be expressed as 
\begin{equation}
    \begin{aligned}
    P &= \sum_i p_i \\
    &= \sum_i \rho(x_i^t).
    \end{aligned}
\end{equation}

\subsection{Fractional existence}
\label{sec:fractional_existence}

We are ultimately interested in the inclusion of atomic existences as a continuous variable, but 
will first present a more general adaptation of the message-passing framework where each node 
has an associated existence. The inclusion of node existence as a continuous variable requires upholding a few 
conditions: 
\begin{itemize}
    \item \textbf{Embedding equivalence}: The presence of a node with null existence should have no influence on the embeddings of other nodes.
    \item \textbf{Embedding continuity}: Node embeddings should change continuously with the existence of neighbouring nodes.
    \item \textbf{Readout equivalence}: The presence of a node with null existence should have no influence on any readouts.
    \item \textbf{Readout continuity}: Readout predictions should change continuously with the existence of nodes.
\end{itemize}

These conditions may be enforced with only minor modifications to the message-passing 
formulation. If we denote the existence of node $i$ as $q_i$, a continuous number in the interval $[0, 1]$, 
and define $Q(q)$ as a continuous function with boundary conditions $Q(0) = 0$ and $Q(1) = 1$, then message-passing can 
be restated as 
\begin{equation}
    \begin{aligned}
        x_i^k &= \gamma^k\left[ x_i^{k-1}, 
        \underset{j\in\mathcal{N}(i)}{\oplus}
        \  \phi^k\left(x_i^{k-1}, x_j^{k-1}, \mathbf{e}_{ji}\right) \cdot Q(q_j) \right].
    \end{aligned}
    \label{eq:message_passing_existence}
\end{equation}
Here, $Q(q_j)$ ensures that once aggregated messages are scaled by the existence of the sender, which may either be considered 
a modification of the aggregation operator or a modification of the message function. Two common choices of 
aggregation functions are summation and averaging which may be adapted to take existence into account like so
\begin{equation}
    \begin{aligned}        
    m_i &= \sum_j m_{ji} \rightarrow \sum_j m_{ji} \cdot Q(q_j). \\
    m_i &= \frac{1}{|\mathcal{N}(i)|}\sum_j m_{ji} \rightarrow \frac{1}{\sum_{j \in \mathcal{N}(i)} Q(q_i)}\sum_j m_{ji} \cdot Q(q_j).
    \end{aligned}
\end{equation}
Which ensures that the influence of node $j$ is proportional to its existence, having full influence when $q_j = 1$ and no influence 
when $q_j = 0$. The conditions on readouts can be similarly enforced, 
\begin{equation}
    P = \sum_i \rho(x_i^t) \cdot Q(q_i).
\end{equation}
Where, $Q(q_i)$ now ensures that the contribution of node $i$ to the readout property $P$
is proportional to the existence of node $i$. This enforces the two readout conditions. 
These modifications ensure the stated conditions, and as such enable the prediction of e.g.
total energies of atomic systems where each atom also carries an existence variable $q_i$. 

As stated, all that is required for the function $Q(q)$ is that it is continuous and 
has the proper boundary conditions, however we suspect that choosing a monotonic 
function will be beneficial when the intended use is with gradient-based optimization. Therefore, we 
choose the simple linear function as it fits all the requirements
\begin{equation}
Q(q) = q.
\label{eq:linear_q_func}
\end{equation}
The choice of the $Q(q)$ function controls the behaviour of the learned function
for non-integer node existences. Training the network only requires data with 
integer existence and thus posses no extra complications compared to 
training a regular MLIP. The function $Q(q$) can in fact be changed 
after training which means that fractional existence can be added to 
a pretrained GNN, such as the $u$MLIPs that have become 
widespread in recent years - that implicitly have $Q(q) = 1$. In Appendix \ref{app:fingerprint_derive} we show that 
this formulation allows the derivation of the existence encoded fingerprint proposed by Larsen et. al \cite{larsenMachinelearningenabledOptimizationAtomic2023} 
and in Appendix \ref{app:lennard_jones} how we can derive a pair potential that includes atomic existence.

\subsection{Message Passing for Atomic Systems}
\label{sec:mp_atoms}

The application of message passing for atomic systems requires encoding the 
system as a graph with appropriate choices of initial node and edge embeddings and 
is typically enhanced by a physically coherent choice of message function. 

We choose a relatively simple architecture based on the SchNet \cite{schuttSchNetContinuousfilterConvolutional2017}, but note that 
any architecture that can be cast in the form of \eqref{eq:message_passing} can 
be modified similarly. In SchNet atomic embeddings are initialized 
based on the atomic numbers $Z_i$,
\begin{equation}
x_i^0 = x^0(Z_i)
\end{equation}
where $x^0$ is a learnable embedding function that transforms the atomic number into a $v$-dimensisonal vector. 
Edge embeddings are created by expanding the distances between nodes in a basis, e.g. consisting of Gaussians, such that 
\begin{equation}
e_{ij} = \mathrm{RBF}(r_{ij})
\label{eq:rbf_basis}
\end{equation}
The number $b$ of different basis functions determine the dimensionality of $e_{ij}$, that is $\mathrm{RBF}: \mathbb{R}^1 \rightarrow \mathbb{R}^b$. Previously we used bold face 
to indicate the possible vectorial properties of the edge embeddings but SchNet only uses the radial edge embeddings, so 
we omit that now. 

The SchNet message function can now be stated as
\begin{equation}
    m^k_{ji} = \phi_2^k\left(W^k(e_{ji}) \odot \phi_1^k(x_j^{k-1})\right) \cdot f_c(r_{ij}),
\end{equation}
where $W^k: \mathbb{R}^b\rightarrow \mathbb{R}^v$ is a learnable function that 
produces a radial filter, that is multiplied elementwise with the output from $\phi^k_1: \mathbb{R}^v \rightarrow \mathbb{R}^v$. 
All three of these $W^k$, $\phi_1^k$ and $\phi_2^k: \mathbb{R}^v \rightarrow \mathbb{R}^v$ are simple feed-forward neural networks, holding the majority of the trainable 
parameters of the model. Finally, $f_c(r_{ji}) = \frac{1}{2}(\cos(\frac{\pi r_{ji}}{r_c})+1)$ is a cutoff function, ensuring that messages 
go smoothly to zero at the cutoff radius $r_c$. Conceptually, the function $Q(q)$ plays the same role for existences as the cutoff function does for distances, both ensuring 
that the messages are propagated correctly. 
We employ sum aggregation taking existence into account,
\begin{equation}
    m_{i} = \sum_{j\in\mathcal{N}(i)} m_{ji} \cdot q_j.
\end{equation}
And finally the update function $\gamma^k: \mathbb{R}^v \times \mathbb{R}^v \rightarrow \mathbb{R}^v$ introduces a skipped connection 
\begin{equation}
    \begin{aligned}
    x^k_i &= \gamma(x_i^{k-1}, m_i) \\
    &= x_i^{k-1} + m_i    
    \end{aligned}
\end{equation}
We will use this architecture for the prediction of total energies using the common 
anzats of decomposing the total energy into atomic contributions computed using the 
atom embeddings after $t$ iterations of message-passing. Accounting for existence this 
results in
\begin{equation}
    E = \sum_i \epsilon(x_i^t) \cdot q_i,
\end{equation}
where $\epsilon$ is a learnable function mapping atomic embeddings to local atomic energies.
This allows for the computation of gradients of the total energy with respect to both the 
Cartesian coordinates and the existence - both employing automatic differentiation.
To penalize short bonds between atoms with non-zero existence we also add a repulsive pair potential, see Appendix \ref{app:lennard_jones} for details.

While this work does not employ uncertainty estimates, we note that the existence-augmented
GNN can be combined with standard approaches such as ensembles, dropout, or a dedicated uncertainty head. 
Because the message-passing framework is continuous in the existence variable $q$, 
both predicted energies and their associated uncertainties would vary smoothly with $q$. 
In the case of ensembles, each model interpolates its own predictions at fractional 
$q$, and the ensemble variance is then computed, yielding continuous uncertainties.
We emphasize that such uncertainty estimation is beyond the scope of the present 
study but is fully compatible with the proposed framework.

\subsection{Gibbs free energy}

The Gibbs free energy $\Delta G$ of a surface, neglecting vibrational and entropic contributions, may be stated as
\begin{equation}
    \Delta G(x) = E(x) - E_{\mathrm{slab}} - \sum_i^N (\Delta \mu_{Z_i} + \varepsilon_{Z_i}).
\end{equation}
Where $E$ is the total energy, $E_{slab}$ is the energy of the bare slab, $N$ is the total number of atoms and $\Delta \mu_Z$ and $\varepsilon_Z$ 
are the chemical potentials and reference energies of species Z. Ordinarily this expression cannot be 
optimized using local optimization techniques employing gradients. Instead, a typical 
approach is to search for the global minimum for each stoichiometry under consideration, 
which may employ gradient-based optimization of $E(x)$ and then calculating the Gibbs free energy 
as a post-processing step. 

To incorporate existence into this expression we note that the adapted 
message-passing framework we presented in Section \ref{sec:graphs_mp} allows us to write 
the total energy as a function of both Cartesian coordinates and existences, so we can write the 
Gibbs free energy as a function of these as well
\begin{equation}
    \Delta G(x, q) = E(x, q) - E_{\mathrm{slab}} - \sum_i q_i (\Delta \mu_Z + \varepsilon_Z).
\end{equation}
From this expression we can obtain derivatives of $G(x, q)$ with respect to both Cartesian coordinates $\frac{\partial \Delta G(x, q)}{\partial x_i}$ 
and atomic existences $\frac{\partial \Delta G(x, q)}{\partial q_i}$ that can be employed to minimize the Gibbs free energy of 
atomic configurations. 

\subsection{Grand Canonical Optimization}
\label{sec:grand_canon_opt}

\FigureTwo

To avoid having to impose bounds on the optimization of the existence $q$ we 
introduce the variable $\chi$ that is related to $q$ through a sigmoid transform
\begin{equation}
    q(\chi) = \frac{1}{1+\exp(-\chi)},
\end{equation}
so that the free variable is $\chi$. 

Having atomic existences allows us to introduce additional biases,
for example we can add an extra term in addition to the Gibbs free energy that drives the optimization towards a prescribed 
total existence of a particular element using a harmonic expression
\begin{equation}
    l(q) = k\left( \left[\sum_i q_i I(Z_i=Z) \right] - N_Z \right)^2.
    \label{eq:count_harmonic_bias}
\end{equation}
Here $k$ is the familiar strength of the harmonic potential, $I$ is the indicator function, $Z$ is the atomic species the
potential acts on and $N_Z$ is the location of the minimum e.g. the amount of existence of that species that minimizes the term. 
This enables more control of the optimization procedure for species that are not 
suitably described by a chemical potential. 

Figure \ref{fig:opt_frames} illustrates our optimization, the procedure starts with an initial configuration 
with a large number $>100$ of atoms with random positions and existences, as seen in the first frame of Figure \ref{fig:opt_frames}. 
We then simultaneously optimize both the positions and existence variables 
using a variant of the FIRE algorithm \cite{bitzekStructuralRelaxationMade2006}, until the maximum derivative of the objective 
with respect to the existence of any atom is less than a predefined threshold. Once the 
threshold is reached we set the existence of all atoms with an existence above 0.9 
to 1 and all those lower than 0.9 to 0, occurring between the two final frames in Figure \ref{fig:opt_frames}, and continue the optimization with fixed existence until 
the Cartesian forces are below a predefined threshold. To compensate for the anisotropy between the two types of variables $\{x_i\}$ and $\{q_i\}$, our 
FIRE implementation works on each coordinate type separately. The distributions of existences for each frame are shown in Appendix \ref{app:existence_dist}.

\subsection{Dataset \& Training}
\label{sec:training_dataset}

In the following sections we will employ the existence adapted 
graph neural network we have presented. Specifically
will utilize it for a copper-oxide surface, to this end we 
train the network on a dataset of copper-oxides in different cells 
than the one we are interested in. In particular the dataset 
consists of 
\begin{itemize}
\item Low energy Cu(111) $c(8\times4)$ configurations of various stoichiometries from our recent work \cite{christiansenDmodelCorrectionFoundation2025}.
\item Subsampled local optimization trajectories of Cu(111) $c(8\times4)$ of various stoichiometries.
\item Subsampled local optimization trajectories of randomly generated configurations of oxygen on $3 \times 3$ surfaces for the (100), (110) and (111) planes.
\item Configurations of the "29" Cu(111) surface oxide.
\end{itemize}
The full datasets consists of 58534 configurations, we label the dataset using the Orb 
$u$MLIP, particularly the orb-v3-conservative-inf-omat variant. The full set 
is split into train, validation and test sets according to an 80/10/10 split.
Distributions of key properties for this dataset are shown in Appendix \ref{app:dataset}

The architecture is the one described in Section \ref{sec:mp_atoms} using
64-dimensions for both atomic and for Gaussian edge embeddings and three iterations of message-passing 
each followed by layer normalization of the atomic embeddings. Atomic graphs are constructed 
with a cutoff of $5$ Å. We emphasize that training the network is perfectly identical to training a 
regular neural network MLIP, the addition of the existence coordinate has no influence as in all training configurations
all atoms have unity existence. We train towards both the energy and the forces with 0.1/0.9 
weight between the two. We train for 400000 gradient steps and select the model 
with the lowest error on the validation set. The learning rate is initialized to 
$10^{-4}$ and is adjusted if the validation error stagnates. Our implementation 
uses PyTorch, PyTorch lightning and PyTorch geometric \cite{paszkePyTorchImperativeStyle2019, feyFastGraphRepresentation2019, falconPyTorchLightning2025}. 
Training the model took $\sim 2$ hours on an Nvidia A100 GPU. 

\section{Results}

\subsection{Existence conditions}

\FigureThree

As noted in Section \ref{sec:fractional_existence} our inclusion of existence is 
guided by four conditions. To verify that our methodology 
correctly upholds these conditions we perform a series of small 
experiments.  

First, to verify embedding equivalence we present Figure \ref{fig:existence_conditions}(a-b) where two dimensions of the atomic 
embeddings for a \ch{CuO3} cluster and for a \ch{CuO4} cluster where the additional 
atom has null existence, are shown to be equivalent. Quantitatively we can also report 
that the maximum difference in any of the embeddings computed for this comparison is on the 
order of $10^{-7}$, an acceptable error that can be attributed to 
numerical differences. As the existence increases we require the embeddings to continuously rotate and scale, 
which can be observed by comparing Figure \ref{fig:existence_conditions}(b-d) where the existence of 
the extra atom is changed from 0 through 0.5 to 1.
Lastly Figure \ref{fig:existence_conditions}(e) deals with the two 
conditions on readouts. In this case our readout is the total 
energy, which we require to be continuous with respect to the existence and 
at the extremes be equivalent to the same configuration with and 
without the inclusion of the atom with varying existence. As seen 
from the line starting as black in the top left corner the 
energy decreases nearly linearly and eventually reaches the exact value 
of the CuO$_4$ structure, indicated by the dashed red line, in the lower right corner. 
The nearly linear behaviour of the energy is caused by the linearity of our choice of $Q(q)$, Eq. \eqref{eq:linear_q_func}, and the network 
recognizing that \ch{CuO4} has a lower total energy than \ch{CuO3}. This linear behaviour is curiously
similar to the behaviour of DFT with fractional particle number \cite{perdewDensityFunctionalTheoryFractional1982}.

\FigureFour

\begin{table}
    \centering
    \begin{tabular}{|l|l|}
        \hline
        $\Delta \mu_\mathrm{O}$ [eV] & $\{-2 -1.5 -1.25, -1, -0.5\}$ \\ \hline
        $N_{Cu}$ & $\{2, 3, 4, 5, 6\}$ \\ \hline
    \end{tabular}
    \caption{Table of the values of $\Delta \mu_\mathrm{O}$ and $N_{Cu}$ resulting in 25 combinations of 
    the two parameters.}
    \label{table:parameters}
\end{table}

\section{$c(6\times2)$  copper-oxide}

We employ the described grand canonical optimization method to a $c(6\times2)$-copper oxide surface at every 
combination of the chemical potentials $\Delta \mu_\mathrm{O}$ and bias on the number of copper $N_{Cu}$ listed 
in Table \ref{table:parameters}. For each combination we generate 250 
configurations resulting in a total of 6250 structures. The chemical potentials are chosen to 
cover a wide range of experimentally relevant values. Similarly, we introduce a bias toward 2-6 
Cu atoms in order to discourage multilayer structures. 

As the starting point for the grand canonical relaxation we randomly place 75 atoms of each species with random 
initial existence and follow the procedure described in Section \ref{sec:grand_canon_opt} while keeping the positions and existences of the slab 
atoms fixed. The time pr. optimization step is comparable to using the model for relaxation with respect to just the Cartesian coordinates and 
can be done on either CPU or GPU depending on the available resources. In this case we used 4 CPU cores with the time pr. configuration generated 
ranging between 5-10 minutes depending on the number of gradient steps necessary to reach the convergence conditions. 
We set the reference energy of oxygen to half the energy of an \ch{O2}-molecule, the reference of copper is calculated 
based on a bulk copper cell, both using the model and $\Delta \mu_\mathrm{Cu} = 0$ eV. In Figure \ref{fig:cu_6x2_configs} we show 
the most favourable configuration found at the different chemical potentials investigated and note that the one
identified for $\Delta\mu_O = -1.25$ eV is identical to the stable phase previously reported in the literature \cite{feidenhanslOxygenChemisorptionCu1101990}.

\FigureFive

To investigate our method's ability to produce configurations that are relevant at a desired chemical potential we plot, in Figure \ref{fig:gibbs_dist}(a-c), the distribution of 
Gibbs free energies for configurations found with the method targeting the desired chemical potential and compare that to the distribution of 
Gibbs free energies of configurations found targeting other chemical potentials. In all cases the method produces configurations that are more relevant when 
targeting the desired chemical potential than when not, as evidenced by those distributions being concentrated at lower Gibbs free energies. 
We also present the distribution of the number of oxygen in the generated configurations at different chemical potentials of oxygen, see Figure \ref{fig:gibbs_dist}(d), 
which follows the expectation of less oxygen being present at lower values of $\Delta \mu_\mathrm{O}$. Finally, we investigate
how the number of oxygen correlates with the number of copper at a particular chemical potential of oxygen. 
In Figure \ref{fig:gibbs_dist}(e) we observe that the algorithm covers a range of values for the number of copper, despite the use of Eq. \eqref{eq:count_harmonic_bias} with $k = 5$ eV, 
and tends to allow for the existence of more oxygen when more copper is present. 

For each of the 72 stoichiometries generated by the grand canonical optimization procedure we take the most favourable one according to the model and evaluate its energy using DFT. 
We use the GPAW DFT code with the PBE functional with a 500 eV plane-wave cutoff, a $4 \times 4$ $\mathbf{k}$-point grid for the surface reconstructions
and a $12 \times 12 \times 12$ $\mathbf{k}$-point to calculate the reference energy of copper from a bulk cell and a spin-polarized calculation for \ch{O2} 
to calculate the oxygen reference energy \cite{hjorthlarsenAtomicSimulationEnvironment2017, mortensenGPAWOpenPython2024}. This allows us to calculate a DFT-based Gibbs free energy diagram that we present in Figure \ref{fig:dft_gibbs_diagram}. This again identifies 
the \ch{Cu5O4} as the most stable phase for a range of $\Delta \mu_\mathrm{O}$ in agreement with \cite{feidenhanslOxygenChemisorptionCu1101990}.
It is inarguable that the phase diagrams produced using the model's energy predictions and the DFT energies 
do not match in general. The network was trained using data labelled with Orb, if the Gibbs diagram is calculated 
with Orb it is in general agreement with the DFT diagram - identifying \ch{Cu2O2} and \ch{Cu5O4} as the only two stable oxide phases. 
Therefore, our model's disagreement stems from its own inaccuracies. The model and the surrounding optimization algorithm 
may be considered both a MLIP and a generative model - where inaccuracies in the MLIP will affect the generated 
structures - even so it was able to generate a wide range of relevant configurations. Inaccuracies 
in the MLIP may be amended by additional training data or a more sophisticated message-passing neural network architecture. 
A common workflow for training MLIPs is subsampling of high temperature molecular 
dynamics to ensure the potential has been exposed to those regions of the PES, 
this may be extended with grand canonical optimization trajectories as to further strain the potential.

\FigureSix

\section{Conclusion}

We have presented a formulation of the message-passing framework employed by state-of-the-art 
graph neural network MLIPs that incorporates fractional atomic existence as an additional variable. This addition 
enables gradient-based grand canonical optimization, which we demonstrate for a copper surface oxide 
where we are able to selectively produce relevant configurations at a chosen chemical 
potential. Our methodology imposes no constraints on the network architecture nor introduces 
additional requirements on the training data, it can in fact even be added to preexisting models without affecting 
their predictions for configurations where all atoms fully exist. The resulting model is both a MLIP and 
can be considered a generative model when combined with an optimization procedure. 

While we exploit the presented method's ability to get derivatives of the total energy with respect to the atomic existence 
the architectural changes we propose may find uses elsewhere. One such case is for generative models, a number of which 
rely on similar architecture as those employed for MLIPs. For example, diffusion models, where current methods prescribe 
the desired number of atoms in order to keep the number of dimensions consistent, could be augmented to 
include the atomic existence giving these models more freedom \cite{xieCrystalDiffusionVariational2022, hoogeboomEquivariantDiffusionMolecule2022, lyngbyDatadrivenDiscovery2D2022, ronneGenerativeDiffusionModel2024}. 

\section{Data availability}
Datasets are available on Zenodo at \zenodo \ covering the employed training dataset with orb-v3-conservative-inf-omat labels and the generated
Cu(110) $c(6\times 2)$ reconstruction configurations. \newline
Source code available from Gitlab at \gitlab \ under an MIT license. 

\section{Acknowledgements}
We acknowledge support from VILLUM FONDEN through Investigator grant, project 
no. 16562, and by the Danish National Research Foundation through the Center of 
Excellence “InterCat” (Grant agreement no: DNRF150). We thank Prof. Johannes Margraf 
for pointing out the similarity to DFT with fractional particle number. 

\bibliographystyle{unsrt}
\bibliography{references}

\appendix
\section{Existence fingerprint derivation}
\label{app:fingerprint_derive}

Here we show that our formulation encompasses the inclusion of existence through modification of a fingerprint descriptor $\rho(r)$ as proposed by Larsen et. al \cite{larsenMachinelearningenabledOptimizationAtomic2023}.
To this end the fingerprint may be considered a global graph property, one way of framing it in this way is to choose the initial atomic embeddings to be null vectors and to consider the 
message function to be 
\begin{equation}
    m_{ji}(r) = \frac{1}{r_{ij}^2} f_c(r_{ij}) \exp{\left(-\frac{|r-r_{ij}|^2}{2\delta^2_R}\right)}
\end{equation}
Where $f_c(r_{ij})$ is a smooth cutoff function that decays to zero at a specified cutoff radius, and we have chosen $Q(q) = q$. 
With sum aggregation accounting for existence and an identity update function this means that node embeddings become 
\begin{equation}
    x_i(r) = \sum_{j \in \mathcal{N}(i)} q_j \frac{1}{r_{ij}^2} f_c(r_{ij})  \exp{\left(-\frac{|r-r_{ij}|^2}{2\delta^2_R}\right)}
\end{equation}
And a readout layer that sums over all nodes leads to 
\begin{equation}
    \begin{aligned}
    \rho(r) &= \sum_i q_i \sum_{j \in \mathcal{N}(i)} q_j \frac{1}{r_{ij}^2} f_c(r_{ij})  \exp{\left(-\frac{|r-r_{ij}|^2}{2\delta^2_R}\right)} \\
            &= \sum_{\substack{i,j \\ i \neq j}} q_i q_j \frac{1}{r_{ij}^2} f_c(r_{ij}) \exp{\left(-\frac{|r-r_{ij}|^2}{2\delta^2_R}\right)}
    \end{aligned}
\end{equation}
Where in the second line the neighbourhood function $\mathcal{N}(i)$ is chosen as all atoms in a structure and locality is solely 
encoded by the cutoff function. The second line is the modified fingerprint expression used by Larsen et al. In practice $r$ 
would be included as a hyperparameter, in the same fashion as the number of basis functions in Eq. \eqref{eq:rbf_basis}. 

\section{Pair potential with existence}
\label{app:lennard_jones}

As an example of a pair potential we consider the Lennard Jones potential. To compute the Lennard Jones energy as a graph property we can start by again considering initial atomic embeddings of zero, 
and messages correspond to half the bond energy 
\begin{equation}
    m_{ji} = 2 \epsilon \left[ \left(\frac{\sigma}{r_{ji}}\right)^{12} - \left(\frac{\sigma}{r_{ji}}\right)^{6} \right].
\end{equation}
Which with sum aggregation accounting for existence and an identity update function leads the atomic embedding to be 
\begin{equation}
    x_i = \sum_j 2 q_j \epsilon \left[ \left(\frac{\sigma}{r_{ji}}\right)^{12} - \left(\frac{\sigma}{r_{ji}}\right)^{6} \right].
\end{equation}
And the total energy can be achieved using a summation readout layer
\begin{equation}
    E = \sum_{\substack{i,j \\ i \neq j}} 2 q_iq_j \epsilon \left[ \left(\frac{\sigma}{r_{ji}}\right)^{12} - \left(\frac{\sigma}{r_{ji}}\right)^{6} \right].
\end{equation}
The same derivation holds for a general pair potential where the bond energy is $\epsilon(r_{ij})$, leading to
\begin{equation}
    E = \sum_{\substack{i,j \\ i \neq j}} \frac{1}{2}q_i q_j \epsilon(r_{ij}).
\end{equation}
For the repulsive pair potential we employ the bond energy is given by 
\begin{equation}
    \epsilon(r_{ij}) = 
    \begin{cases}
    & \alpha \left(1 - \left(\frac{r_{ij}}{r_{min}}\right)^2\right)^3, \quad r_{ij} < r_{\mathrm{min}} \\
    & 0, \quad r_{ij} \geq r_{\mathrm{min}}
    \end{cases}
\end{equation}
Where $\alpha = 100$ is the maximum value at $r_{ij} = 0$ and $r_{\text{min}}$ is $\frac{5}{4}$ the sum of covalent radii of the 
involved atomic species.

\section{Existence distribution}
\label{app:existence_dist}
\begin{figure*}
\centering
\includegraphics[]{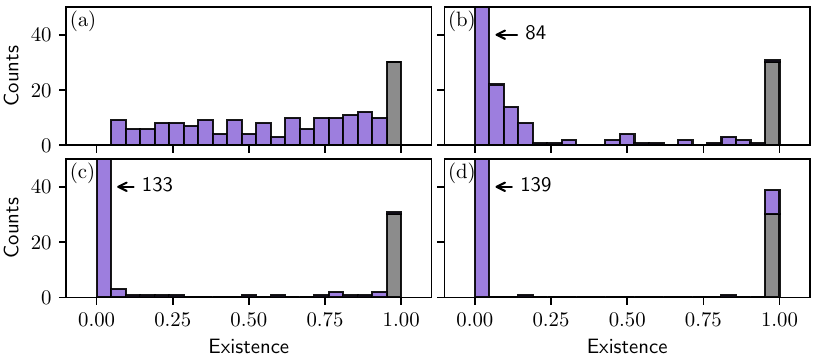}
\caption{Existence distributions for the first four frames shown in Figure~\ref{fig:opt_frames}. The 
distribution for the atoms optimized by the gradient-based grand canonical optimization 
procedure is depicted in purple and for the static template slab in gray.}
\label{fig:existence_dist}
\end{figure*}

Figure \ref{fig:existence_dist} shows the distribution of existence values for the 
optimization trajectory depicted in Figure \ref{fig:opt_frames}.
\section{Dataset distribution}
\label{app:dataset}

\begin{figure*}
\centering
\includegraphics[]{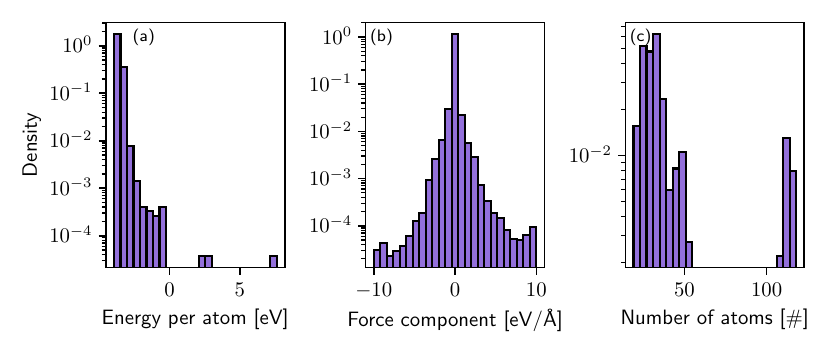}
\caption{Distributions of (a) energy pr. atom, (b) force components and (c) number of atoms 
in the copper-oxide training dataset. The energy and force labels are from the orb-v3-conservative-inf-omat model.
The range of the distribution of force components covers 99.95\% of the dataset with the remaining 0.05\% having $|F|>10$ eV/Å.}
\label{fig:data_dist}
\end{figure*}

Figure \ref{fig:data_dist} shows the distributions of key properties of the training data set. 

\end{document}